\documentclass{article}

     \usepackage[final,nonatbib]{neurips_2021}

\usepackage[utf8]{inputenc} 
\usepackage[T1]{fontenc}    
\usepackage{hyperref}       
\usepackage{url}            
\usepackage{booktabs}       
\usepackage{amsfonts}       
\usepackage{nicefrac}       
\usepackage{microtype}      
\usepackage{xcolor}         
\usepackage{wrapfig}
\usepackage{subcaption}
\usepackage{float}
\usepackage{graphicx}

\usepackage[utf8]{inputenc}
\usepackage{bbm}
\usepackage{hyperref}
\usepackage{amsthm,textcomp}
\usepackage{float}
\usepackage{indentfirst}
\usepackage{cleveref}

\hypersetup{
    colorlinks=true,
    linkcolor=blue,
    filecolor=blue,      
    urlcolor=blue,
}

\title{Advanced Methods for Connectome-Based Predictive Modeling of Human Intelligence: A Novel Approach Based on Individual Differences in Cortical Topography} 

\author{%
  Evan D. Anderson\\
  Neuroscience Program\\
  University of Illinois\\
  Urbana, IL \\
   \And
  Ramsey Wilcox\\
  Department of Psychology\\
  University of Illinois\\
  Urbana, IL \\   
  \And
  Anuj Nayak\\
  Department of Electrical \& Computer Engineering\\
  University of Illinois\\
  Urbana, IL \\
    \And
  Christopher Zwilling\\
  Beckman Institute\\
  University of Illinois\\
  Urbana, IL \\
    \And
  Pablo Robles-Granda\\
  Department of Computer Science\\
  University of Illinois\\
  Urbana, IL \\    
  \And
  Been Kim\\
  Google Brain\\
  Mountan View, CA \\
      \And
  Lav R. Varshney\\
  Department of Electrical \& Computer Engineering\\
  University of Illinois\\
  Urbana, IL \\
      \And
  Aron K. Barbey\\
  Department of Psychology\\
  University of Illinois\\
  Urbana, IL \\
}
\date{September 2021}
\begin{document}

\maketitle

\begin{abstract}
Individual differences in human intelligence can be modeled and predicted from in vivo neurobiological connectivity. Many established modeling frameworks for predicting intelligence, however, discard higher-order information about individual differences in brain network topology, and show only moderate performance when generalized to make predictions in out-of-sample subjects. In this paper, we propose that connectome-based predictive modeling, a common predictive modeling framework for neuroscience data, can be productively modified to incorporate information about brain network topology and individual differences via the incorporation of bagged decision trees and the network based statistic. These modifications produce a novel predictive modeling framework that leverages individual differences in cortical tractography to generate accurate regression predictions of intelligence scores. Network topology-based feature selection provides for natively interpretable networks as input features, increasing the model's explainability. Investigating the proposed modeling framework's efficacy, we find that advanced connectome-based predictive modeling generates neuroscience predictions that account for a significantly greater proportion of variance in general intelligence scores than previously established methods, advancing our scientific understanding of the network architecture that underlies human intelligence.

\end{abstract}

\section{Predicting Individual Differences in General Intelligence}

Cognitive neuroscience research has begun to turn greater attention to individual differences in neurobiology and cognition \cite{mcfarland}, motivating the development of research methods that more directly address individual differences in neurosceicne data. Integrating cognitive neuroscience and computer science methods and perspectives together promises to advance the explanability and replicability of scientific results \cite{hofman}, and will be of particular importance for advancing research into the neruosceince of inter-individual differences \cite{fellous}. In this paper, we propose a novel predictive modeling framework that modifies connectome-based predictive modeling (CPM) \cite{finn,shen}, an existing computational approach for predicting behavioral data from neuroscience data, and demonstrate the method's efficacy for predicting intelligence using individual differences in cortical topography.

\subsection{Existing Computational Cognitive Neuroscience Approaches for Predicting Intelligence}

Intelligence is a central trait underpinning individual differences in cognitive ability \cite{caemmerer}. While the neurobiological basis of intelligence has long been identified with the architecture of specific brain networks \cite{jung}, recent work has begun to establish that that systemwide brain network topology and dynamics are critical sources of the broad individual differences observed in cognitive ability \cite{aron}.

Several previous methods have been applied to model and predict human intelligence from neuroscience data. Connectome-based predictive modeling (CPM) has previously been deployed to predict fluid intelligence scores \cite{finn}, showing that CPM's predictions could account for 25\% of the variance observed across actual fluid intelligence. Other approaches, such as cortical hyperalignment, have accounted for up to 39\% of the variance in general intelligence scores on the basis of single brain region's connectivity, and on average account for 27\% of variance in general intelligence scores when using whole-brain connectivity \cite{feilong}. The efficiency of weakly-connected edges (extracted from windowed functional connectivity) have been used in a correlation framework to account for 37.5\% of the variance in IQ \cite{emiliano}. Elastic net models are able to account for 20\% of the variance in intelligence scores, on the basis of distributed network of resting state connections that span the functional connectome \cite{dubois}. Each method has desirable properties---use of strong connections, use of weak connections, native interpretability, predictive accuracy---that our advanced modeling framework will attempt to incorporate.

Modeling and generating predictions from individual differences in functional connectivity remains an important goal for neuroscience \cite{gabrieli}, requiring interpretable AI methods that model individual variability in neuroscience data \cite{fellous}. Here, we investigated the feasibility of modifying CPM to create an advanced predictive modeling framework that incorporates information about network topology and individual differences. One previously unremarked feature of CPM is that it ablates brain network topology during feature selection (see \cite{shen}). As neuroscience evidence suggests that individual differences in network topology are important for explaining cognitive abilities, we incorporated a natively interpretable network-based feature selection into our advanced CPM to extract individual differences in cortical topology.

\subsection{Connectome-Based Predictive Modeling}

We first deployed standard connectome-based predictive modeling to quantify the baseline variance in general intelligence in our sample accounted for by standard CPM \cite{shen}. CPM filters brain regions using mass-unvariate statistical thresholding against the behavioral outcome, maximizing the proportion of true positives edges included in the model. These features are then summarized (i.e., added) to create a single value per subject, which train a linear model between aggregate neuroscience data and behavioral outcome. The procedure is performed under cross-validation to generate behavioral predictions for each subject. Our baseline implementation of CPM incorporated edges with both positive and negative relationships to intelligence simultaneously, aggregating edges by sign and then combining those two values to produce a single value per subject.

\subsection{Network-Based Statistic}

To include interpretable information about network topology, our advanced CPM replaces mass-univariate feature selection with a network based statistic (NBS) model \cite{zalesky}. NBS uses permutation testing to assess the relationship between individual network edges and a behavioral outcome, explicitly considering network topology while deploying an interpretable linear model to identify edges that form a significant and topologically connected network. Although isolated functional edges contains information about the BOLD timeseries of two incident regions, the mesoscale and macroscale topology of the larger networks cannot be easily inferred or reconstructed from many such aggregated edge weights. Deploying NBS to perform feature selection allows our advanced CPM framework to train on features that not only relate significantly to intelligence, but also form a connected network, explicitly including network topology information into the modeling process.

\subsection{Bootstrap Aggregation}

We further modified CPM to attend to individual differences by eliminating the feature summarization step, instead training with disaggregated features (i.e., topologically connected functional edges) using bagged random forests \cite{khaled}. Bagged random forests have have several desirable properties, including a low number of model parameters and resistance to overfitting in the training population (as individual trees are uncorrelated). Here, we selected bagged random forests for two primary reasons. First, disaggregating edge strengths preserves individual difference information about network topology, and bagged random forests are capable of training on these individual edges and producing a regression output. Second, each tree is bootstrapped by oversampling from a subpopulation in the data (ie, bootstrap resampling), allowing for an ensemble of models that are trained (and overfit) with variability unique to subpopulation of individuals. The resulting decision trees do not model variability common to the whole sample, but instead will overfit to variance unique to the bootstrap replicate, characterizing sources of individual difference unique to that subpopulation. Aggregating the predictions of many uncorrelated decision trees can lead to good performance for generalizing from highly variable data (such as functional connectivity data, see \cite{turner}).

\subsection{Data Acquisition}

We acquired a large ($N=297$) dataset of 10-minute resting state EPI scans using 3 Tesla MRI, and processed them using reproducible methods for analysis of neuroimaging data (via ICA-AROMA denoising \cite{aroma} through FMRIPREP \cite{fmriprep} and xcpEngine \cite{xcp}). Diffusion tensor imaging data was also acquired in $N=288$ subjects on the half-shell and processed using FSL's FDT and bedpostx for probabalistic diffusion tensor tractography \cite{fsl}. Imaging data were parcellated using a multimodal 360-region atlas of cortical grey matter \cite{glasser}. Additionally, we administered a comprehensive battery of neuropsychological tests and deployed structural equation modeling to assess individual differences in general, fluid, and crystallized intelligence.

\subsection{Results and Discussion}

\begin{figure}[H]
    \centering
    \title{Network-Based Statistic identifies fMRI and DTI networks that explain differences in general intelligence}
    \begin{subfigure}[H]{0.97\textwidth}
    \includegraphics{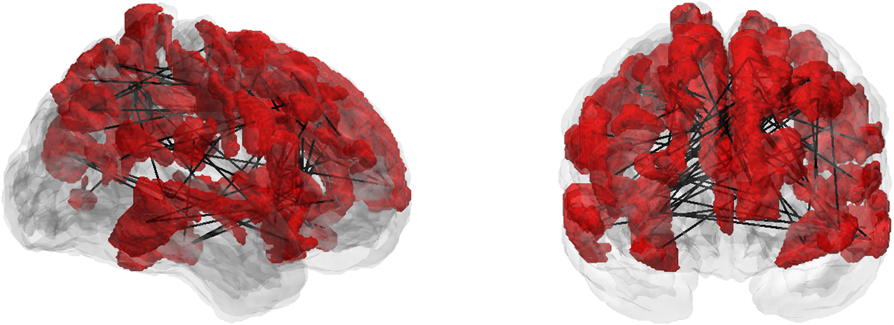}
    \caption{NBS identifies a network of weak connections from resting state connectivity data at $t<-3.5$ and $p<.01$ (finding that functional edge strength was inversely related with intelligence).}
    \end{subfigure}\hfill%
    \hspace{1cm}
    \begin{subfigure}[H]{0.97\textwidth}
    \includegraphics{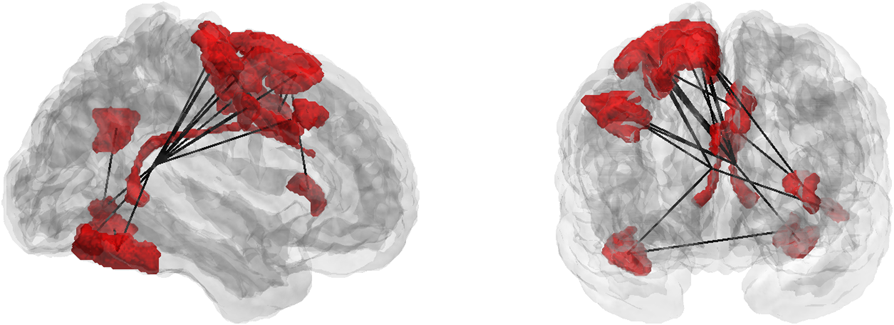}
    \caption{NBS identifies a network of dense structural connections from diffusion tensor tractography that are positively associated with intelligence at $t>3.5$ and $p<.01$.}
    \end{subfigure}\hfill%
    \caption{Network-based statistic provides natively interpretable feature selection in functional (a) and structural (b) connectivity data.}
    \label{fig:NBS}
\end{figure}

\begin{figure}[H]
    \centering
    \begin{subfigure}[t]{0.45\textwidth}
    \includegraphics[width=6.5cm, height=4.7cm]{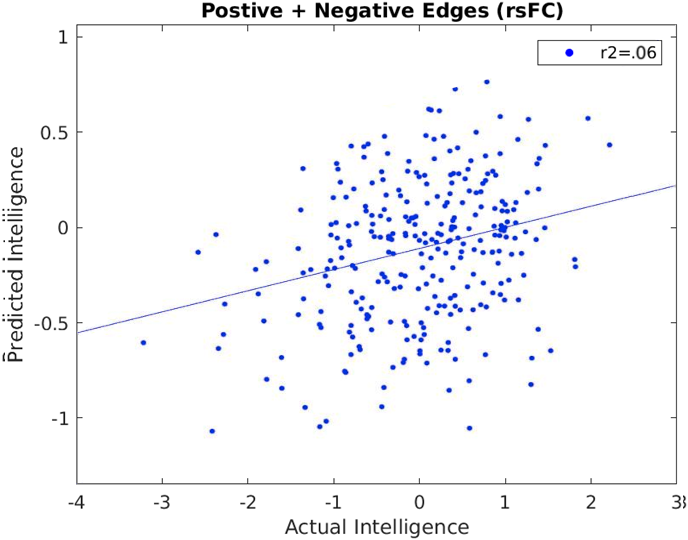}
    \caption{Connectome-based predictive modeling results predict intelligence at $r=0.25$ and $p=.02$.}
    \label{fig:bagga}
    \end{subfigure}\hfill%
    \hspace{1cm}
    \begin{subfigure}[t]{0.45\textwidth}
    \includegraphics[width=6.5cm, height=4.7cm]{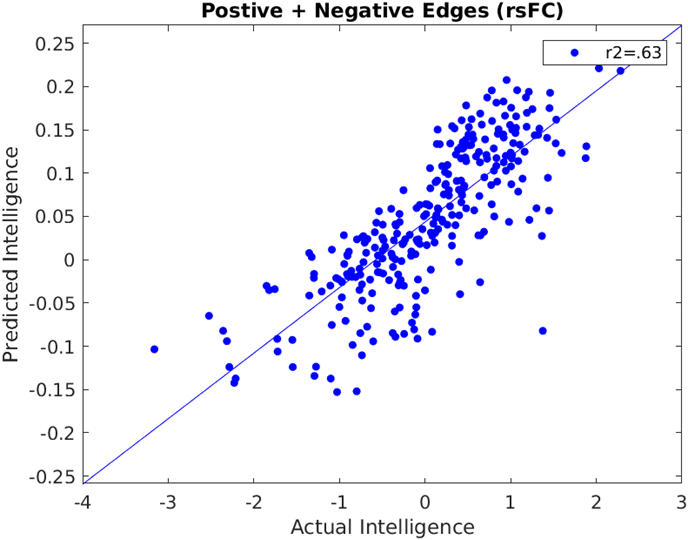}
    \caption{Advanced CPM using resting state connectivity predicts intelligence at $r=0.79$ and $p<.01$.}
    \label{fig:baggb}
    \end{subfigure}\hfill%
    \begin{subfigure}[t]{0.45\textwidth}
    \includegraphics[width=6.5cm, height=4.7cm]{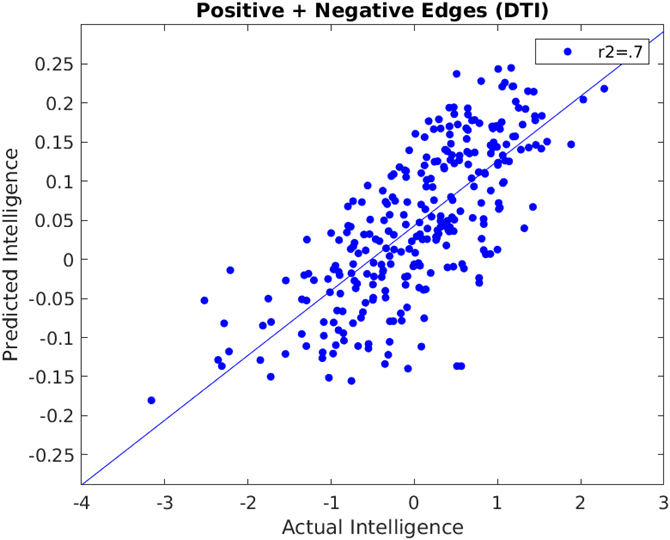}
    \caption{Advanced CPM using structural connectivity predicts intelligence at $r=0.84$ and $p<.01$.}
    \label{fig:baggc}
    \end{subfigure}\hfill%
    \begin{subfigure}[t]{0.45\textwidth}
    \includegraphics[width=6.5cm, height=4.7cm]{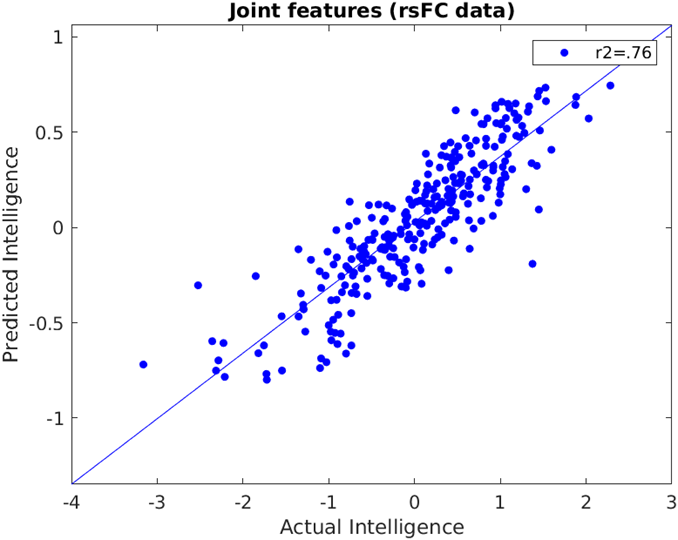}
    \caption{Advanced CPM with multimodal selection predicts intelligence at $r=0.87$ and $p<.01$.}
    \label{fig:baggd}
    \end{subfigure}\hfill%
    \caption{Predictive model performance for CPM (a) and advanced CPM (b,c,d).}
\end{figure}

CPM predictive accuracy (fig. \ref{fig:bagga}) was notably lower than in a previous report \cite{finn}, possibly explained by our smaller sample size, our more robust motion artifact removal \cite{parkes}, and the broader set of cognitive operations entailed by our dependent variable $g$. Network-based statistic identified both functional and structural networks reliably associated with general intelligence scores (fig. \ref{fig:NBS}), and these network edges produced reliable N-fold predictions of intelligence when combined with bagged random forests, accounting for 63--70\% of the variance between subjects (\cref{fig:baggb,fig:baggc}). The range of predictions was notably narrow---this may be due in part to the sign consistency of feature-selected edges imposed by NBS (either all positive or all negative) as a consequence of t-thresholding.

CPM trains using edges both positively and inversely associated with intelligence, unlike the individual sign-consistent networks identified by NBS from DTI and fMRI data. Finally, to induce this bidirectional property in advanced CPM, we feature selected the union of network edges identified from fMRI and DTI data. This multimodal network selection retained weak functional connections inversely associated with intelligence, and added functional edges for which underlying white matter connectivity was positively associated with intelligence. Incorporating both strong and weak functional edges improved prediction spread and accounted for 76\% of the variance in intelligence (fig. \ref{fig:baggd}).

In all cases, variance explained by advanced CPM exceeded previous findings in the literature \cite{finn,feilong,emiliano,dubois}. Our results show that advanced CPM generates reliable behavioral predictions when trained using individual difference in natively interpretable cortical networks. Our results suggest that both strong and weak functional edges contain important predictive signals, and that multimodal feature selection produces the highest accuracy predictions, suggesting that individual differences in global network topology are critical in producing intelligence. These findings motivate future work using Shapley values to more precisely interpret individual differences in topological feature importance. Our findings highlight the utility of incorporating intrepretable information about network topology into the study of individual differences in cognition, and further advance our scientific understanding of the network architecture underlying human intelligence.

\nocite{*}
\bibliographystyle{IEEEtran}
\bibliography{bibfile}

\end{document}